\documentclass[useAMS,usenatbib]{mnras}
\usepackage{graphicx}
\usepackage{amsmath}
\usepackage{amssymb}
\graphicspath{{./image/}}

\def\aj{AJ}
\def\araa{ARA\&A}
\def\apj{ApJ}
\def\apjl{ApJL}
\def\apjs{ApJS}
\def\ao{Appl.~Opt.}
\def\aap{A\&A}
\def\mnras{MNRAS}
\def\na{New A}
\def\physrep{Phys.~Rep.}

\def\FIX{\textcolor{black}}
\def\FIXII{\textcolor{black}}

\title[Optimal identification of H\scalebox{0.7}{II} regions]{Optimal identification of H\scalebox{0.7}{II} regions during reionization in 21-cm observations}

\author[Giri et al.]{
Sambit K. Giri,\thanks{E-mail: sambit.giri@astro.su.se}
Garrelt Mellema,
and Raghunath Ghara
\\
Department of Astronomy and Oskar Klein Centre, Stockholm University, AlbaNova, SE-106 91 Stockholm, Sweden\\
}

\date{Accepted 2018 June 29. Received 2018 June 25; in original form 2018 January 18}

\pubyear{2018}
\begin{document}
\label{firstpage}
\pagerange{\pageref{firstpage}--\pageref{lastpage}}
\maketitle

\begin{abstract}
The ability of the future \FIX{low frequency component of the Square Kilometre Array radio telescope (SKA-Low)} to produce tomographic images of the redshifted 21-cm signal will enable direct studies of the evolution of the sizes and shapes of ionized regions during the Epoch of Reionization. However, a reliable identification of ionized regions in noisy interferometric data is not trivial. Here, we introduce an image processing method known as superpixels for this purpose. We compare this method with two other previously proposed ones, one relying on a chosen threshold and the other employing automatic threshold determination using the K-Means algorithm. We use a correlation test and compare power spectra and bubble size distributions to show that the superpixels method provides a better identification of ionized regions, especially in the case of noisy data. We also describe some possible additional applications of the superpixel method, namely the derivation of the ionization history and constraints on the source properties in specific regions.
\end{abstract}




\section{Introduction}
The observation of the redshifted 21-cm signal, which is produced by the spin-flip transition of neutral hydrogen, is expected to revolutionize the study of the era when the Universe reionized, commonly known as the Epoch of Reionization (EoR). Various indirect observational probes, such as 
the study of high-redshift quasars, the cosmic microwave background radiation and Ly$\alpha$ emitting galaxies have 
constrained the completion of reionization to a redshift close to 6 \citep[e.g.][]{2015ApJ...802L..19R, 2015MNRAS.454L..76M}. However, there is still a large degree of uncertainty regarding the duration of this event and the nature of the sources of ionizing radiation. 
%
As the redshifted 21-cm signal directly probes the neutral hydrogen, it will allow us to measure how and when it disappeared and thus should provide much better constraints on the answers to these questions \citep[e.g.][]{2006PhR...433..181F,2010ARA&A..48..127M,2012RPPh...75h6901P}.

It is, however, very challenging to detect the 21-cm signal from the EoR as it is rather weak and contaminated by foreground signals which are orders of magnitude stronger. A new generation of low frequency radio telescopes, such as the Giant Metrewave Radio Telescope \citep[GMRT; e.g.][]{2011MNRAS.413.1174P}, the Low Frequency Array \citep[LOFAR; e.g.][]{2010MNRAS.405.2492H}, the Murchison Widefield Array \citep[MWA; e.g.][]{2009IEEEP..97.1497L}, the Precision Array for Probing the Epoch of Reionization \citep[PAPER; e.g.][]{2010AJ....139.1468P}, have been attempting to detect this signal. Even though they have not yet been successful, some useful upper limits on the 21-cm power spectrum have been derived, for example by PAPER \citep{2015ApJ...801...51J} and LOFAR \citep{2017ApJ...838...65P}.

All the above radio telescopes, as well as the new HERA telescope, currently under construction \citep{deboer2017hydrogen}, aim to study the 21-cm signal from the EoR using statistical estimators such as the variance and power spectrum. However, the \FIX{low frequency component of the future Square Kilometre Array \citep[SKA-Low;][]{2013ExA....36..235M}}, will have the additional capability to produce images of the 21-cm signal. Observing over a range of frequencies the EoR can be studied as a sequence of images, each from a different redshift. The study of these images at different frequencies (or redshifts) is known as 21-cm tomography \citep[e.g.][]{furlanetto200421}. Tomographic data sets will allow us to study the evolution of the signal and hence the progress of reionization.

To study tomographic images using metrics such as size, shape, orientation etc, we have to establish which regions represent what.
Since the 21-cm signal is produced by neutral hydrogen it might seem straightforward to identify ionized regions or `bubbles' in these images as they would have zero signal. However, several factors make it highly non-trivial to find ionized regions this way. First of all the signal depends not only on the density of neutral hydrogen but also on its spin temperature and when this is equal to the temperature of the Cosmic Microwave Background (CMB) this will also result in a zero signal. However, the expectation is that at late times the high spin temperature limit applies, making the signal only dependent on the neutral hydrogen density \citep{2012RPPh...75h6901P}. Even under these conditions, identifying ionized regions is not easy as finite resolution will complicate disentangling fluctuations in the neutral hydrogen fraction and density. The presence of noise will add another level of uncertainty. All of this is made worse by the fact that interferometric images do not have an absolute flux calibration as they only measure fluctuations, implying that we do not know the zero level in our images.

One proposed method to identify ionized regions in 21-cm data sets is the matched filter technique \citep[e.g.][]{2013ApJ...767...68M,2016JApA...37...27D}. The filter used is based on theoretical expectations of the shapes of ionized regions and is applied directly to the raw visibilities observed by the radio interferometer. The technique works well for noisy data and as long as ionized bubbles are isolated. However, after they start overlapping, their shapes become very complicated and it becomes impossible to know a priori what filtering kernel to use. 

Identification of ionized regions in tomographic images was part of the studies of the ionized bubble size distributions from simulated 21-cm tomographic data in \citet{2017MNRAS.471.1936K} and \citet{giri2017bubble}. The former used a simple constant threshold value and the latter explored the use of a varying threshold value based on the probability distribution function (PDF) of the measured signal. Both of these papers pointed out limitations of these methods but also identified regimes in which they seem to work well. However, it is clear that there is a need for more reliable methods.

The research field of computer vision has been very active in developing techniques to identify meaningful objects in images. This has led to a new field of study known as image segmentation. An image is partitioned into disjoint subsets (or segments) whereupon subsets of interest can be studied. The earliest inspiration for image segmentation came from gestalt theory \citep{wertheimer1944gestalt}, which describes the relation between meaningful patterns and grouping laws such as similarity, proximity and continuity. Recent overviews of the field of image segmentation can be found in \citet{nagabhushana2005computer} and \citet{szeliski2010computer}. 

\FIX{In this study we show how the image segmentation method known as over-segmentation with superpixels can be used to identify ionized regions in tomographic 21-cm images.} We describe it as well as the methods used in \citet{2017MNRAS.471.1936K} and \citet{giri2017bubble} in terms of the theory of image segmentation and compare the new method to these previously proposed ones. As no real data is available we base our exploration, just as the previous studies mentioned above did, on mock observations constructed from reionization simulations and a telescope model.

The structure of the paper is as follows. \FIX{In the next section we describe our procedure to create mock observations that should mimic the results from the Phase 1 version of SKA-Low, henceforth SKA1-Low.} In Section 3, we present the different image segmentation methods for identifying ionized regions. Section 4 describes the application of the new method to simulated 21-cm data sets. The results of a comparison between the different segmentation methods are found in Section 5, which is followed by a section describing a few applications made possible by a reliable identification of ionized regions. We summarize the conclusions of our study in Section 7.

\section{Simulated 21-cm signal} 
To test the different bubble identification methods we use mock 21-cm observations created from a reionization simulation. We employ the latest design for the antenna distribution and noise characteristics of SKA1-Low to generate data sets which have realistic noise levels and resolution. We assume that any residual foreground signals are below the noise level.

\subsection{Numerical simulation}
The cosmological 21-cm signal is calculated from the results of a fully numerical reionization simulation. These simulations consist of two stages. First, the evolution of the matter distribution is calculated using the N-body code \textsc{CUBEP$^3$M} \citep{2013MNRAS.436..540H}. Next, the ionization structure of the matter is calculated with \textsc{C$^2$-RAY} \citep{2006NewA...11..374M}, a radiative transfer simulation code. Our simulation methodology has been described in more detail in previous papers \citep[e.g.][]{2006MNRAS.372..679M, 2006MNRAS.369.1625I, 2012MNRAS.424.1877D}.

The simulation used here is the LB1 simulation from \citet{2016MNRAS.456.3011D}. It has a comoving volume of (349Mpc)$^3$ distributed over a mesh consisting of $250^3$ cells. 
\FIX{The sources of ionizing photons are located in dark matter haloes that are more massive than $10^9$~M$_\odot$. Such haloes can develop star forming regions through atomic cooling and are also fairly insensitive to radiative feedback \citep[see][and references therein]{2018MNRAS.473...38S}. These haloes are assumed to release ionizing photons at a rate of 1.7 photons per baryon every $10^7$~years.} The cosmological parameters used here are $\Omega_\mathrm{m}$=0.27, $\Omega_k$=0, $\Omega_\mathrm{b}$=0.044, $h=0.7$, $n_\mathrm{s}=0.96$ and $\sigma_8$=0.8. These values are consistent with the results from \textit{Wilkinson Microwave Anisotropy Probe} (WMAP) \citep{2011ApJS..192...18K} and Planck \citep{2016A&A...594A..13P}.

\subsection{Redshifted 21-cm signal}
\label{sec:21-cm_sig}

The differential brightness temperature of the observed 21-cm signal is given as \citep[e.g.,][]{2013ExA....36..235M},
\begin{eqnarray}
\delta T_\mathrm{b} \approx 27 x_\mathrm{HI} (1 + \delta)\left( \frac{1+z}{10} \right)^\frac{1}{2}
\left( 1 -\frac{T_\mathrm{CMB}}{T_\mathrm{s}} \right)\nonumber\\
\left(\frac{\Omega_\mathrm{b}}{0.044}\frac{h}{0.7}\right)
\left(\frac{\Omega_\mathrm{m}}{0.27} \right)^{-\frac{1}{2}} 
\left(\frac{1-Y_\mathrm{p}}{1-0.248}\right)
\mathrm{mK}\ ,
\label{eq:dTb}
\end{eqnarray}
where $x_\mathrm{HI}$ and $\delta$ are the local fraction of neutral hydrogen and the density fluctuation respectively. $T_\mathrm{s}$ is the local excitation temperature of the two spin states of the neutral hydrogen, known as the spin temperature. $T_\mathrm{CMB}$ is the CMB temperature at redshift $z$ and $Y_\mathrm{p}$ is the primordial helium abundance. 

From Equation~\ref{eq:dTb}, it is clear that there will be no signal when $T_\mathrm{CMB} \approx T_\mathrm{s}$. It is expected that during the EoR the spin temperature decoupled from $T_\mathrm{CMB}$ and approached the gas temperature due to the Wouthuysen-Field effect \citep{1997ApJ...475..429M}. This makes the 21-cm signal observable. When the gas temperature is below $T_\mathrm{CMB}$,  the signal is seen in absorption and when it is above, it is seen in emission. In this study, we assume the high spin temperature limit, $T_\mathrm{s}\gg$ $T_\mathrm{CMB}$, in which case the signal becomes independent of the value of $T_\mathrm{s}$. This is generally considered to be a valid assumption for the later stages of reionization as even relatively low levels of X-ray radiation can raise the gas temperature above the CMB temperature \citep[e.g.][]{2007MNRAS.376.1680P,2012RPPh...75h6901P}. In this case regions with $\delta T_\mathrm{b}=0$ can be easily \FIX{associated with} ionized regions.

Using Equation~\ref{eq:dTb} we construct three-dimensional 21-cm data sets from the neutral fraction and density fields produced for a given redshift $z$ by \textsc{C$^2$-RAY} and \textsc{CUBEP$^3$M} respectively. This 3D data set is taken to represent an image cube where one of the axes is taken to be the frequency axis and the other two represent position in the sky. We disregard any evolution of the signal along the frequency direction, i.e.\ we use coeval cubes, not light cone cubes. \FIX{We also do not change the frequency dependent telescope parameters over the data cube, i.e.\ we use the noise and resolution corresponding to the redshift of the coeval cube.} 

The intrinsic resolution of this data set is $\Delta x=1.39$ comoving Mpc which corresponds to angular scales
\begin{equation}
\Delta \theta = \frac{\Delta x}{D_\mathrm{c}(z)}\,,
\end{equation}
where $D_\mathrm{c}(z)$ is the comoving distance to redshift $z$, and frequency scales
\begin{equation}
\label{eq:freq_resolution}
\Delta \nu = \frac{\nu_0 H(z) \Delta x}{c(1+z)^2}\,,
\end{equation}
where $\nu_0$ is the rest frequency of the 21-cm line, $H(z)$ the Hubble parameter at redshift $z$ and $c$ the speed of light. For $z=7$, these equations give values $\Delta \theta=0.535$ arcmin and $\Delta \nu=0.085$~MHz.

\begin{figure*}
  \centering
  \includegraphics[width=0.9\textwidth]{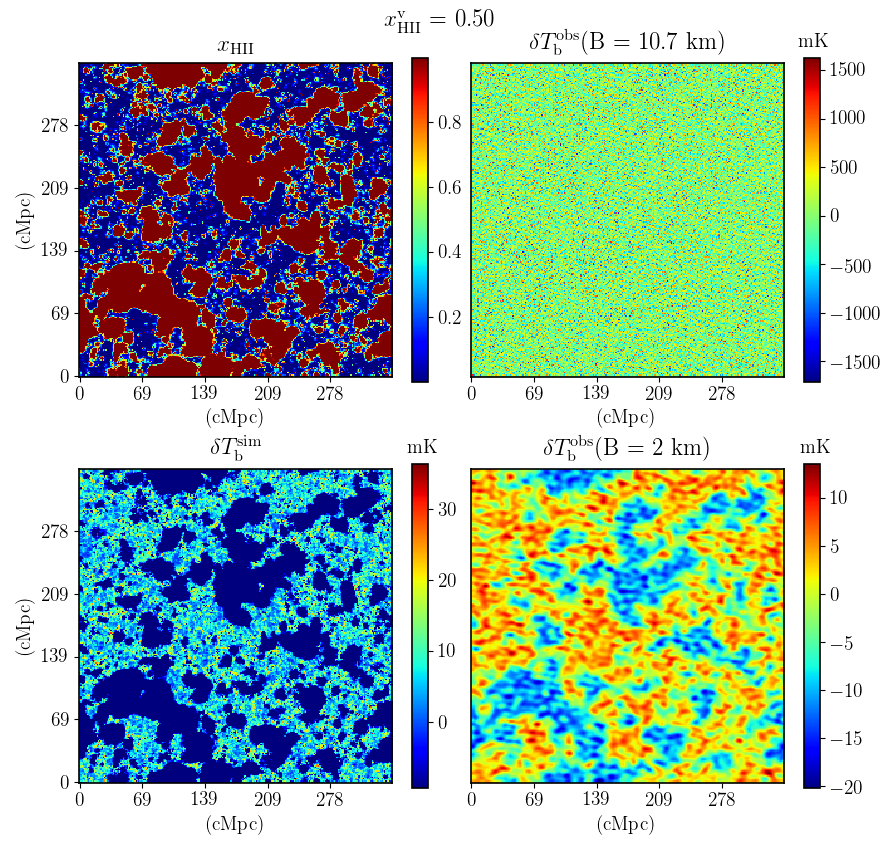}
  \caption{Slices from our reionization simulation for $z=6.905$ \FIX{($x^\mathrm{v}_\mathrm{HII}=0.50$)}. \textit{Top left}: The ionization fraction at the resolution of the simulation. \textit{Bottom left}: The \FIX{intrinsic} 21-cm signal at the resolution of the simulation. We can clearly see the features of the ionized regions here. \textit{Top right}: The \FIX{observed} 21-cm signal with noise at a resolution corresponding to a maximum baseline of 10.7~km (matching the resolution of the simulation); the noise has an rms value of 393~mK, which is much higher than the rms of the 21-cm signal. \textit{Bottom right}: The \FIX{observed} 21-cm signal with noise smoothed with a kernel corresponding to a maximum baseline of 2 km. The features of the ionized regions become visible at this resolution. The rms of this data set is 5.98 mK. \FIX{The rms of our modelled telescope noise at this resolution is 2.82 mK.}}
  \label{fig:mock_obs}
\end{figure*}

\subsection{Characteristics of the interferometer}
\label{base_ska}
In this study we only consider SKA1-low, the only low frequency radio interferometer sensitive enough to produce tomographic 21-cm data sets of the EoR \citep{2015aska.confE..10M}. \FIX{As the configuration and the antennas for SKA1-low are still being developed, our telescope parameters should be considered as indicative for what SKA1-Low can achieve. We use a configuration which has 512 antenna stations\footnote{\FIX{The configuration is described in document SKA-TEL-SKO-0000557 Rev 1 and the positions of the individual stations in document SKA-TEL-SKO-0000422 Rev 2, both retrievable at https://astronomers.skatelescope.org/documents/}} each with a diameter of 35 meter. Among these, 224 non-overlapping stations are randomly distributed to form a compact core of diameter 1 km. The rest of the stations are grouped into 48 clusters, each of which contains six randomly placed stations. These clusters are distributed in a three-armed spiral with a 35 km radius.} The \textit{uv} coverage of the SKA1-low used in this study is generated assuming a 4 hour observation towards a region with declination -30$^\circ$. The resulting \textit{uv} coverage is very similar to the one presented in \citet{2017MNRAS.464.2234G}.


\subsection{Noise map}
\label{sec:noise_map}

Each pair of stations in SKA1-Low will record noise along with the visibilities. This noise is uncorrelated and can be represented by Gaussian random numbers. The mean of these numbers is taken to be zero and the standard deviation is given by the following equation \citep[e.g.,][]{2017MNRAS.464.2234G},
\begin{equation}
\sigma = \frac{\sqrt[]{2}k_B T_\mathrm{sys}}{\epsilon A_\mathrm{D} \sqrt[]{\Delta \nu t_\mathrm{int}}} \ .
\end{equation}
In the above equation, $k_\mathrm{B}$ and $T_\mathrm{sys}$ 
are the Boltzmann constant and the system 
temperature of the telescope. $\Delta \nu$ and $t_\mathrm{int}$ are the frequency resolution of the observed data and the integration time. 
\FIX{The former is calculated using Equation~\ref{eq:freq_resolution} taking $\Delta x$ to be the resolution of our simulation, i.e.\ 1.39~comoving Mpc.} 
$A_\mathrm{D}$ is the physical area of each station and $\epsilon$ an efficiency factor which is given as follows 
\begin{equation}
\epsilon = \left\{\begin{matrix}
1 \ \ \ \ \ , \ \nu \leq \nu_\mathrm{c},\\ 
\left(\frac{\nu_\mathrm{c}}{\nu}\right)^2, \ \nu > \nu_\mathrm{c} \ .
\end{matrix}\right.
\end{equation}
This factor captures the fact that the effective station area decreases for higher frequencies and has a second order dependence on the frequency. The detailed steps to convert the noise in the visibilities to that in that SKA1-Low images are given in more detail in \citet[][]{2017MNRAS.464.2234G}. Table~\ref{tab:telescope_param} lists the values of the telescope properties used to calculate the noise. 

\begin{table}
	\centering
	\caption{The parameters used in this study to model the telescope properties.}
	\label{tab:telescope_param}
	\begin{tabular}{lccccc} 
		\hline
		Parameters & Values \\
		\hline
        Observation time ($t_\mathrm{int}$)& 1000 h  \\
        System temperature ($T_\mathrm{sys}$) & $60 (\frac{\nu}{300\mathrm{MHz}})^{-2.55}$ K  \\
		Effective collecting area ($A_\mathrm{D}$) & 962 $\mathrm{m}^2$  \\
        Critical frequency ($\nu_\mathrm{c}$) & 110 MHz \\
		\hline
	\end{tabular}
\end{table}

\subsection{Signal map}
\label{sec:signal_map}
We construct the 21-cm signal cube as described in Section~\ref{sec:21-cm_sig}. \FIX{The bottom left panel of Fig.~\ref{fig:mock_obs} shows one slice from such a cube, denoted as $\delta T_\mathrm{b}^\mathrm{sim}$. For comparison, we also show the values of the ionization fraction for the same slice (top left panel). We Fourier transform this 21-cm data set and sample it with the \textit{uv} coverage calculated from the SKA1-Low array configuration. Next, we add the system noise to this visibility cube and transform back to the image domain. The resulting image, $\delta T_\mathrm{b}^\mathrm{obs}$, is representative of what SKA1-Low would produce after 1000 hours of integration at the full resolution of our reionization simulation which corresponds at $z=7$ to a maximum baseline $B=10.7$~km. We show it in the upper right panel of Fig.~\ref{fig:mock_obs}. It is clear that at this resolution the 21-cm signal is far below the noise and therefore undetectable.}

However, most of the sensitivity of SKA1-Low is at the shorter baselines which means in practice that the noise level will be lower if one reduces the resolution of the images. In this paper, we choose a resolution corresponding to a maximum baseline of length 2~km. We obtain the image cubes at this resolution by convolving the images in the angular direction with a Gaussian kernel with a FWHM corresponding to the maximum baseline of 2 km and in the frequency direction with a top-hat filter of matching width. For $z=7$, the numerical values for the angular and frequency resolution are 2.9 arcmin and 0.46~MHz which both correspond to 7.6 comoving~Mpc.

The bottom right panel of Fig.~\ref{fig:mock_obs} shows a slice from this $\delta T_\mathrm{b}^\mathrm{obs}$($B=$2~km) image cube. The noise level has been strongly reduced and the larger scale structures in the 21-cm signal have become visible. It is in these $\delta T_\mathrm{b}^\mathrm{obs}$($B=$2~km) images that we will attempt to identify the ionized regions \FIX{in the rest of this paper}. The intensity level of this image shows both negative and positive values as an interferometer such as SKA1-Low does not measure the absolute signal, but only fluctuations on the sky, causing the average of each image to be 0~mK.

Below we often consider three stages of reionization, early, middle and late, characterised by average ionization fractions by volume \FIX{$x^\mathrm{v}_\mathrm{HII}$} of 0.2, 0.5 and 0.8. The redshifts and resolution for these three stages in our reionization simulation are listed in Table~\ref{tab:data_param}. The reionization history of the Universe may of course be different from this, we only use this simulation to test our bubble identification methods as it represents a possible 21-cm signal with the typical patchiness seen in most cases.

\begin{table*}
	\centering
	\caption{The specification of the 21-cm data cubes used, including the angular, frequency and spatial resolution for a maximum baseline of 2~km.}
	\label{tab:data_param}
	\begin{tabular}{lccccc} 
		\hline
		$x^\mathrm{v}_\mathrm{HII}$ & z & $\nu_\mathrm{obs}$(MHz) & $\Delta \theta$ (arcmin) & $\Delta \nu$ (MHz) & $\Delta x$ (cMpc)\\
		\hline
		0.20 & 7.570 & 167 & 3.11 & 0.44 & 8.3 \\
        0.50 & 6.905 & 180 & 2.87 & 0.47 & 7.4 \\
        0.80 & 6.549 & 188 & 2.74 & 0.48 & 7.0 \\
		\hline
	\end{tabular}
\end{table*}

\section{Bubble Segmentation}
\label{sec:bub_seg}
Most previous works dealing with statistical studies of ionized bubbles identified these regions from ionization fraction data cubes using a fixed threshold value, typically 0.5 \citep[e.g.,][]{2011MNRAS.413.1353F,2016MNRAS.461.3361L}. 
However, the choice of the threshold value biases the bubble statistics. An example of this can be seen in fig.~3 of \citet{2011MNRAS.413.1353F}, where the bubble size statistics for the friends-of-friends method is shown for different threshold values. The curves are affected significantly by the choice. If we do not select the threshold properly, then a robust study of the evolution of the bubble sizes will be difficult.

The problem of defining a threshold becomes even harder when we want to study the sizes of ionized bubbles from 21-cm observations. We previously studied how well bubble size distributions can be extracted from 21-cm data in \citet{giri2017bubble}. There we showed that this is possible, even at the typical resolution expected from the future SKA1-Low observations. However, this did require the use of a more sophisticated threshold selection method, namely one based on the probability distribution function (PDF) of the 21-cm signal.

In this study, we will resort to a specific area of image processing, known as image segmentation, to identify the regions of interest (ROI) in the simulated 21-cm observations. Image segmentation methods can be broadly divided into three categories, namely intensity-based, discontinuity-based and region-based algorithms \citep{glasbey1995image}. The discontinuity-based methods work best when there are steep gradients in the data, as is for example the case for the ionization fraction fields from simulations. However, the observed 21-cm signal will be affected by density fluctuations, noise and limited resolution of the telescope which will make these gradients less clear. Therefore, we will not explore any discontinuity-based methods but limit ourselves to intensity-based and region-based methods. In this section, we introduce the different segmentation algorithms that we use to identify ionized regions in the 21-cm observation.



\subsection{Intensity-based methods}
\label{sec:glob_thres}
Intensity-based image segmentation methods determine a global threshold value from the values of the resolution elements (henceforth `pixels') of the data set. Their advantage is that they are generally straightforward and computationally inexpensive \citep{sezgin2004survey}. However, being based on the values of individual pixels, they will be more sensitive to noise as this may push the values of individual pixels below or above the threshold value, leading to a incorrect segmentation. This can also result in the formation of less contiguous ROIs, when noise introduces isolated spots inside the identified regions. \citet{2017MNRAS.471.1936K} observed this effect in their intensity-based analysis of noisy 21-cm data sets.

Some examples of intensity-based methods are the mean method, the histogram method, p-tile thresholding and multispectral thresholding \citep{nagabhushana2005computer}. Here we will use two of them, the mean and histogram methods, both of which have been used before on (simulated) 21-cm data sets.


\subsubsection{Mean method}
The simplest method of selecting a threshold for a 21-cm signal data set is to use the mean of the signal. For interferometric images this implies choosing a constant value of 0~mK. \citet{2017MNRAS.471.1936K} used this method for studying the sizes of ionized regions in 21-cm data sets using the granulometry technique. In their study, they label pixels with a signal less than the mean as ionized and refer to these as `cold spots'. 

The mean threshold tends to divide the data into two equal parts \citep{al2010image_google}. Therefore, it typically works well when the ROI covers about 50\% of the data set. 
In the comparison (Sect.~\ref{sec:comparison}) this method is labelled as {\lq Mean\rq}.

\subsubsection{Histogram method}
\label{sec:PDF_split}
The PDF of the ionization fraction values from EoR simulations shows a bimodal distribution with one peak corresponding to neutral pixels and the other to ionized pixels \citep[see e.g.\ fig. 36 in][]{2006MNRAS.371.1057I}. The same is true for the 21-cm signal provided no contributions from spin temperature fluctuations are present. This makes it possible to select a threshold value that splits the PDF into an ionized and neutral part. 

This type of intensity-based segmentation is known as bimodal histogram thresholding. Its advantage is that it uses the PDF of the data itself to select a threshold without any assumption about the meaning of the absolute values of the pixel intensity values. It can also be trivially expanded to multimodal cases. Its main disadvantage is that it requires the data to have a clear bimodal PDF, which even in the case when the actual values have such a PDF, may be hidden by noise or low resolution. 

In \citet{giri2017bubble}, we used a bimodal histogram method to identify ionized pixels in simulated 21-cm data and then studied bubble size distributions with different methods. We used a one-dimensional K-Means algorithm to find the threshold value separating ionized and neutral pixels\footnote{Strictly speaking the 1D K-Means method does not analyse the PDF but finds clusters in the 1D space of pixel values. However, this effectively makes it a bimodal histogram method.}. See section~3.2 of \citet{giri2017bubble} for a detailed description of our application of the 1D K-Means algorithm on the 21-cm data.

The results in \citet{giri2017bubble} showed that the 1D K-Means method works well as long as the PDF of the 21-cm signal is clearly bimodal. As fig.~3 in that paper shows, the results become unreliable when less than 10 per cent of the universe is ionized. It should also be noted that we there applied the method on noiseless data which is the most favourable case for any intensity-based method. 


For the cases where the PDF is not clearly bimodal but unimodal with an asymmetric tail, it is still possible to define a threshold. The asymmetric tail of the 21-cm PDF is caused by the ionized pixels and evolves into the second peak during the later stages of reionization. To determine a threshold in this case we will use the maximum deviation algorithm \citep{rosin2001unimodal}.

\begin{figure}
  \centering
  \includegraphics[width=0.46\textwidth]{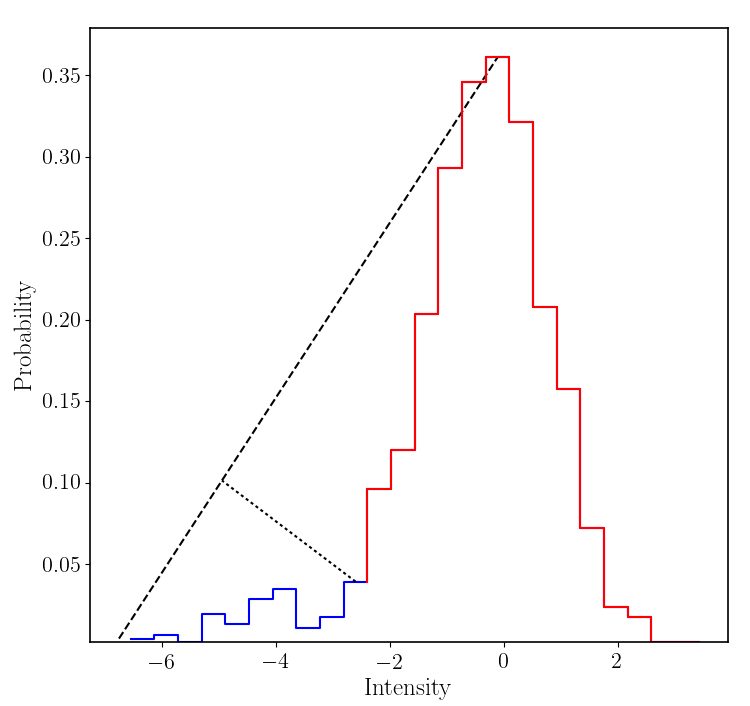}
  \caption{Illustration of the maximum deviation method for choosing a threshold in the case of an asymmetric unimodal PDF. A line is drawn from the peak of the PDF to the end of the tail (dashed-line). The point on the PDF that is farthest from this line gives the threshold to separate out the tail. The dotted line indicates the position of maximum deviation. \FIX{The resulting segmentation of the PDF is indicated in colour.}}
  \label{fig:max_dev_method}
\end{figure}

We explain this algorithm in Fig.~\ref{fig:max_dev_method}. A line is drawn from the peak to the asymmetric end of the distribution. The threshold value chosen corresponds to the point on the asymmetric side of the curve that is farthest away from the line drawn. It should be noted that the number of bins used to produce a PDF will affect the result. Using a large number will make the PDF noisy whereas a smaller number will not capture the intrinsic shape of the PDF. We use Knuth's Bayesian method \citep{2006physics...5197K} to find the optimum number of bins for the given data. 

\begin{figure}
  \centering
  \includegraphics[width=0.46\textwidth]{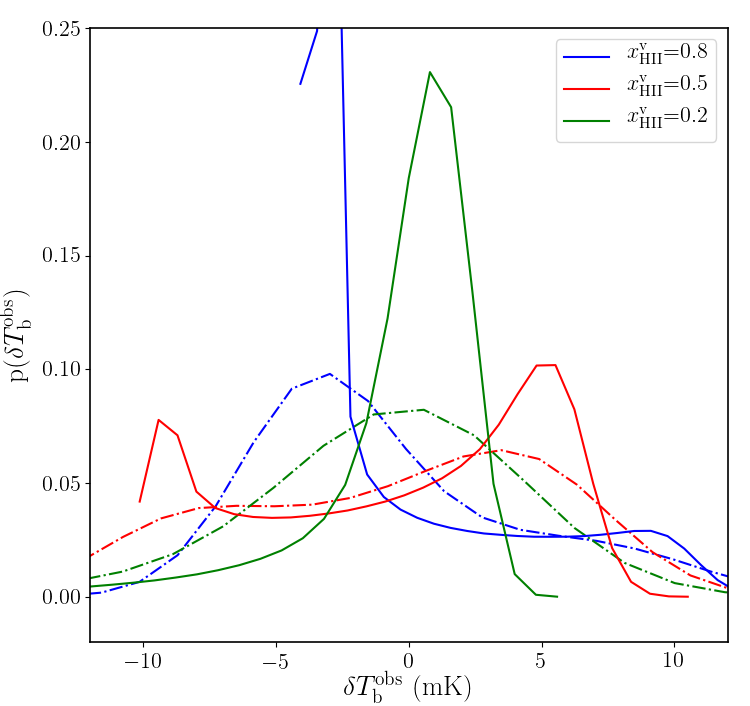}
  \caption{PDFs of the values of the smoothed 21-cm signal for different stages of reionization, colour coded according to the legend. The \FIX{solid} lines shows the PDFs of the smoothed 21-cm signal without noise while the \FIX{dash dotted} lines show the ones with noise. \FIX{Details of the data cubes are given in Table~\ref{tab:data_param}.}}
  \label{fig:pdf_noise_nonoise}
\end{figure}

In Fig.~\ref{fig:pdf_noise_nonoise}, the \FIX{solid lines} show the 21-cm PDFs of the noise free data at global ionization fractions of \FIX{$x^\mathrm{v}_\mathrm{HII}$}=0.2, 0.5 and 0.8. 
The \FIX{dash-dotted} lines show the PDFs for the noisy data. 
The distinct peak in the PDF at a low signal value in the dash-dotted curves for \FIX{$x^\mathrm{v}_\mathrm{HII}$}=0.5 and 0.8 is due to reionization. The pixels which contribute to this peak correspond to ionized regions and a threshold value separating this peak from the rest is easily found. However, this peak is less distinct at early times and/or when the data contains noise. For \FIX{$x^\mathrm{v}_\mathrm{HII}$}=0.2, the noise free PDF shows an asymmetric unimodal distribution. In this case, the maximum deviation method can be used to determine the threshold. However, there is almost no asymmetry in the PDF when noise is added. Therefore, even when adding the maximum deviation method, PDF-based segmentation methods can give poor results during the early stages of reionization.

In Section~\ref{sec:comparison} we will use the combination of the maximum deviation method and the 1D K-Means method and compare the results with those from the other methods. Although it actually consists of a combination of the maximum deviation method and the 1D K-Means method, we will refer to it as {\lq K-Means\rq} during the comparison.

\subsection{Region-based methods}
\label{sec:over_seg}

Region-based methods are useful to identify regions with complex shapes and therefore are ideally suited for identifying ionized bubbles. As they consider the surroundings while segmenting regions, they are also less sensitive to noise. A drawback is that they will typically not be able to identify ROIs which consist of only one or a few pixels. Some examples of region-based methods are the region growth method 
and agglomerative clustering \citep[e.g.][]{nagabhushana2005computer}. In this section we describe the region-based method we selected to use, known as superpixels.

\subsubsection{Superpixels method}
This method partitions an image into many small segments called superpixels. These superpixels are constructed to be as homogeneous as possible, which means that the pixels inside every segment are similar in certain attributes, for example colour intensity. An individual superpixel does not necessarily have a useful meaning as ROIs may be distributed over several superpixels or in other words, the superpixels represent an over-segmentation of the image. Therefore the superpixel method is called an over-segmentation method. However, the superpixels can easily be stitched together or mosaicked using a property of the ROIs to obtain the desired segmentation of the image. We refer to \citet{li2012segmentation}, \citet{zhu2016beyond} and \citet{mehra2016brief} for more details.

Many different algorithms have been developed for over-segmenting images with superpixels \citep[see][]{wei2016superpixel}. In Section~\ref{sec:slic}, we describe the method we use in this study. 
%
%
In Section~\ref{sec:stitching}, we outline how the superpixels can be stitched together to recover the ionized regions in a 21-cm data set. The number of superpixels to use is a free parameter of the method. In Section~\ref{sec:seg_err} we explain how to choose this number.

\paragraph{SLIC}
\label{sec:slic}
The method we use for defining superpixels  is known as Simple Linear Iterative Clustering (or SLIC) which is actually an advanced implementation of the K-Means algorithm. SLIC was proposed by \citet{achanta2012slic} for two-dimensional colour images. Such images form a five-dimensional data set as they have two spatial variables and three colour variables, namely red, green and blue. The higher the number of dimensions, the larger the computational cost \citep[Curse of Dimensionality;][]{bellman2003dynamic}. However, SLIC has an optimized algorithm that works well with higher dimensional data. The performance of SLIC has been compared with other methods in \citet{achanta2012slic} where it was shown that SLIC is not only faster for higher dimensional data but also gives more accurate results. Our 21-cm image cubes form a four-dimensional data set (three spatial dimension and one signal intensity value). Therefore, SLIC is an appropriate choice for over-segmenting our data into superpixels. \FIX{We  have used the SLIC algorithm available in the python package {\sc scikit-image}\footnote{Official website: www.scikit-image.org} \citep{scikit-image}} \FIXII{and implemented the method in {\sc tools21cm}\footnote{{https://github.com/sambit-giri/tools21cm}}, a collection of python routines for producing and analysing 21-cm observables from reionization simulations.}

In order to assign the pixels to a particular superpixel, the method requires a metric which measures the similarity of the pixels. SLIC uses a special distance metric which includes the Euclidean distance between the two pixels together with a difference in the signal intensity between the pixels \citep[][]{achanta2012slic}. We modified this distance measure for our data. We define the distance between two pixels $i$ and $j$ as
\begin{equation}
\noindent D_{ij} = \sqrt[]{\left( \frac{d_{ij}^\mathrm{I}}{M^\mathrm{I}}\right)^2 +\left(\frac{d_{ij}^\mathrm{s}}{M^\mathrm{s}}\right)^2} ,
\label{eq:dist}
\end{equation}
with the (spatial) Euclidean and intensity distances defined as
\begin{eqnarray}
d_{ij}^\mathrm{s} &=& \sqrt[]{(x_i-x_j)^2+(y_i-y_j)^2+(z_i-z_j)^2}, \\ 
\noindent d_{ij}^\mathrm{I} &=& |I_i-I_j|, 
\end{eqnarray}
where $I_i$ represents the intensity of a pixel at position ($x_i,y_i,z_i$). These two distances have different units and therefore need to be normalised. This is done using $M^\mathrm{I}$ and $M^\mathrm{s}$, the maximum intensity and spatial distance in the particular segment. See \citet{achanta2012slic} for more details about the similarity metric. A pseudocode for SLIC is given under Algorithm 1 in \citet{achanta2012slic}.


\begin{figure}
  \centering
  \includegraphics[width=0.48\textwidth]{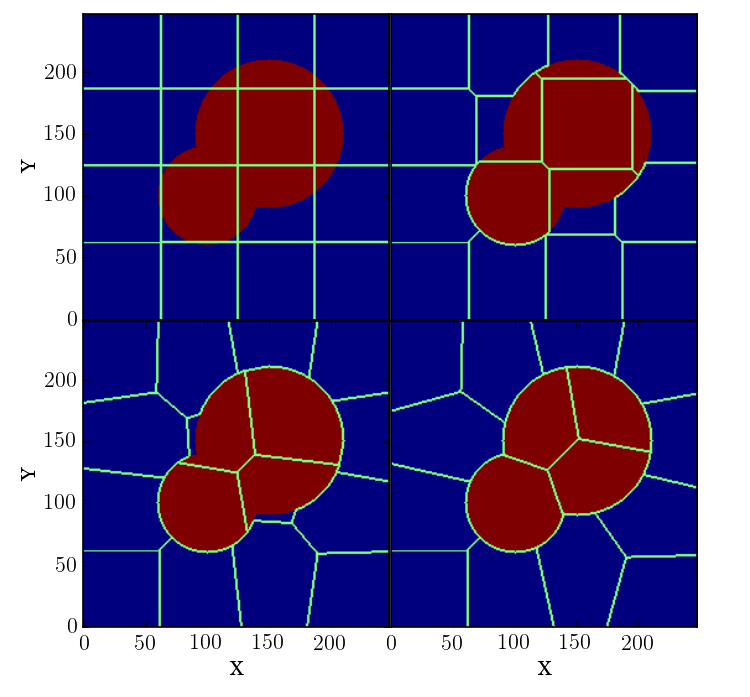}
  \caption{A cartoon showing the progress of the SLIC algorithm for a simple case. The top-left panel shows the initial segmentation. The top-right and bottom-left show the segmentation after 1 and 10 iterations. SLIC has converged in the bottom-right panel.}
  \label{fig:cartoon_slic}
\end{figure}


Figure~\ref{fig:cartoon_slic} gives a pictorial representation of the operation of the SLIC algorithm. We want to identify the red region in the image.
\FIX{We start by choosing $n$ equally spaced pixels and divide the image into equally sized segments with these pixels as their centres (top left panel, $n=16$). We then calculate for each pixel the distance $D$ (Eq.~\ref{eq:dist}) to the different centres and define new segment boundaries so that all pixels within a segment are closer to its centre than to any other centre. This operation is equivalent to a Voronoi tessellation with the distance metric defined in Eq.~\ref{eq:dist}. The new boundaries are shown in top right panel. Next we find the centroids of the new segments and define these to be their new centres. Then we use these new centres to tessellate the image again after which we once more recalculate the centres. The sequence of moving the boundaries and recalculating the centres is repeated until the newly calculated segment centres overlap with the previous ones. This iterative technique is based on the idea of Centroidal Voronoi tessellation \citep[e.g.][]{du2005optimal}. The bottom left panel shows an intermediate iteration where the segments have started to trace the boundaries of the ROI and the bottom right panel displays the converged result.}

In the final over-segmentation in Fig.~\ref{fig:cartoon_slic}, the ROI is distributed over four superpixels. However, SLIC has recovered its edges accurately which is the goal of this first stage. Its capability to trace complex boundaries makes it useful for identifying ionized bubbles in 21-cm image cubes. In the next stage these superpixels are stitched together to recover the ROIs. 


\paragraph{Region stitching}
\label{sec:stitching}
The way in which the superpixels are stitched together defines their meaning. As we opt to identify ionized bubbles, the stitching should be based on a property expected for these regions. For this we return to the PDF of the 21-cm values.

In Section~\ref{sec:PDF_split}, we discussed how the single pixel 21-cm PDF is affected by noise. After having segmented the data set into superpixels, 
\FIX{we construct our superpixel PDF by assigning the 21-cm signal in each superpixel
to be the mean of all the pixels contained within it}. By averaging inside superpixels this distribution will be much less affected by noise. Once the superpixel PDF has been created, we recover the features of reionization seen in the noise free 21-cm PDF. The same algorithms as in the case of the single pixel PDF, see Section~\ref{sec:PDF_split}, can then be used to identify ionized superpixels and to stitch adjacent ones to form the ionized regions.

In Section~\ref{sec:comparison}, we compare this new method, consisting of SLIC over-segmentation and PDF-based stitching, to the two intensity-based methods from Section~\ref{sec:glob_thres}. The new method is labelled as {\lq SLIC\rq}. We would like to point out that even without stitching one can extract useful information from the superpixels. We will discuss some of these other applications in Section~\ref{sec:further}.

\paragraph{Under-segmentation error}
\label{sec:seg_err}

\begin{figure*}
  \centering
  \includegraphics[width=0.8\textwidth]{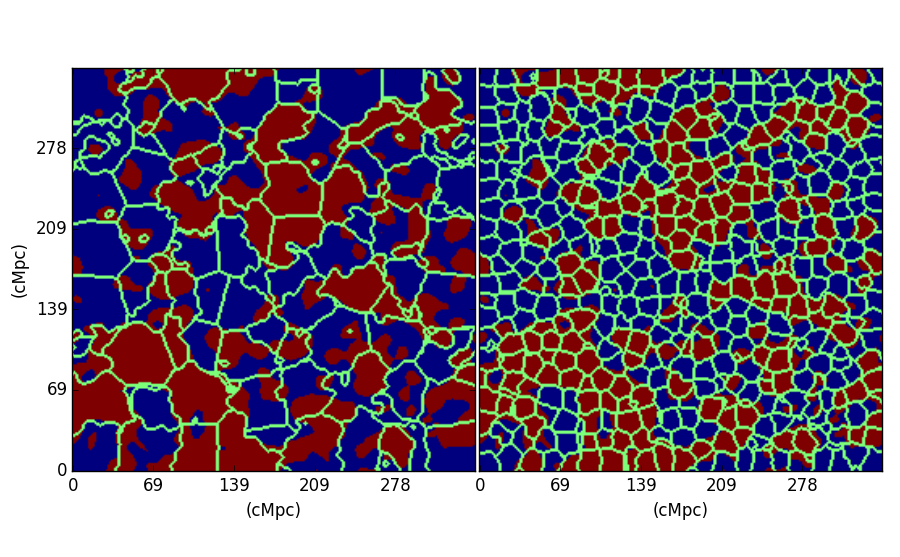}
  \caption{Boundaries of superpixels found for the same slice as shown in the bottom right panel of Fig.~\ref{fig:mock_obs} overlaid on a representation of the ionization fraction in which red is ionized. The left panel shows the result when using 500 segments for the entire data cube and the right panel the result for 5000 segments.}
  \label{fig:seg_underseg}
\end{figure*}

To start the segmentation process we need to specify the number of superpixels to use. If this number is too low, small scale features will not be resolved by the superpixels. In those cases, the data is said to be under-segmented. On the other hand, a large number of superpixels increases the computation time for the segmentation and as the number of superpixels approaches the number of pixels, the result will obviously start to resemble the original image. Therefore, a balanced number of superpixels has to be chosen. For our three dimensional data sets a good first estimate for this number is simply given by 
\begin{equation}
\label{eq:ansatz}
n \approx \frac{L^3}{l_\mathrm{min}^3} \ ,
\end{equation} 
\noindent where $L$ and $l_\mathrm{min}$ are the box size and the smallest scale one is interested in, respectively.

Figure~\ref{fig:seg_underseg} illustrates the impact of the choice of the number of superpixels on our 21-cm data. The left panel shows the superpixel boundaries of a slice from the fiducial data cube (see Fig.~\ref{fig:mock_obs}) segmented into 500 superpixels overlaid on the ionization fraction image, where we reduced the resolution to be the same as in the 21-cm data and take a value of 0.5 and larger to mean ionized. Even though the superpixels correctly delineate the larger regions, they do not trace any of the small scale features. Many superpixels contain a mix of both the ionized and neutral regions, which is undesirable. 
In the right panel, 5000 superpixels have been used and clearly here the boundaries of the ionized regions are traced better. Some superpixels do still contain mixed regions but the overlaps are smaller. Also, the sizes of the superpixels are of course smaller, which makes the contribution to under-segmentation by these superpixels of mixed content substantially less.

For a data set for which the true segmentation is known, the performance of an over-segmentation method can be tested using the so-called under-segmentation error \citep[e.g.][]{levinshtein2009turbopixels,achanta2012slic}, defined as
\begin{equation}
\label{eq:seg_err}
U = \frac{1}{N} \left[ \sum_{i=0,1} \left(\sum_{[s_j|s_j\cap g_i>B]} |s_j|\right) - N \right]\,
\end{equation}
where the $s_j$ are the set of pixels in j$^{\mathrm{th}}$ superpixel. $|s_j|$ gives the number of pixels in that superpixel. $g_i$ represents the set of {\lq ground truth\rq} values of the pixels. The ground truth segmentation is the true segmentation that we are trying to achieve, so ionized and neutral regions in our case. In this study, we derive our ground truth segmentation from the ionization fraction values taken from the simulation. $N$ is the total number of pixels in the data. The quantity $s_j|s_j \cap g_i$ gives the set of superpixels required to cover a ground truth segment $g_i$. A superpixel $s_j$ is considered to be covering the ground truth segment, $g_i$, when at least a fraction $B$ of $s_j$ covers $g_i$. 
We take the value of $B$ to be 0.25.

The lower the under-segmentation error, the better the match to the ground truth. Typically, the under-segmentation error decreases for larger numbers of superpixels. However, as shown by \citet{achanta2012slic}, it will typically stop decreasing beyond a a certain number of superpixels. This number is then the optimal number of superpixels to use. We study this behaviour for our data set in Section~\ref{sec:seg_err_result}.

\FIX{For the real observations we will not know the ground truth. Therefore the choice of the number of superpixels cannot be based on the under-segmentation error.  However, we can still implement the superpixel method by using an iterative process where we choose different values for the number of superpixels. By increasing this number and extracting the ROI we can establish for which number of superpixels the ROI has converged.}

\section{Application of superpixels to 21-cm data}
\label{sec:performance}
\begin{figure*}
  \centering
  \includegraphics[width=1.0\textwidth]{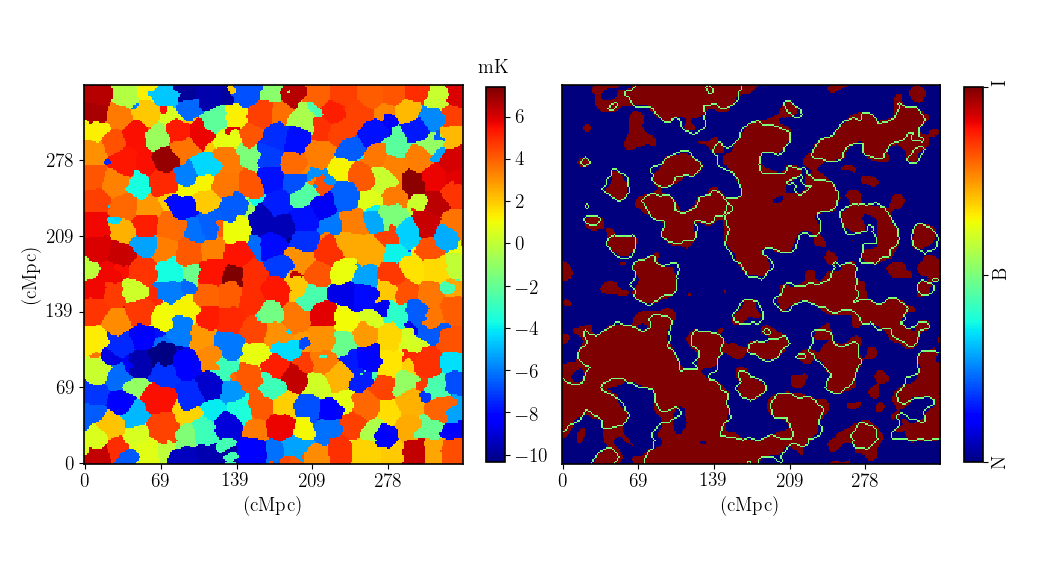}
  \caption{The left panel shows how the slice from the bottom left panel of Fig,~\ref{fig:mock_obs} is segmented into superpixels where the colour coding indicates the mean signal inside each superpixel.  The right panel shows the boundaries of ionized regions identified after stitching overlaid on the ionization fraction slice \FIXII{where red (I) indicates ionized and blue (N) neutral.}}
  \label{fig:superpixeled}
\end{figure*}

Before we compare the results of the different segmentation methods, we first show how the superpixels method performs when used to identify ionized regions from the 21-cm signal. In Fig.~\ref{fig:superpixeled}, we show the superpixels formed from the same slice as shown in the bottom right panel of Fig.~\ref{fig:mock_obs}. The left panel shows the slice with the signal from superpixels. We can see the imprint of the features that we need to identify. In the right panel, the identified boundaries of the ionized regions are shown after stitching. The binary field over which the boundaries are shown is formed from the ionization fraction data set, smoothed to the same resolution as the 21-cm data set, by labelling pixels with value larger than 0.5 as ionized. This reference binary data set will henceforth be referred to as `Ref-Bin'. As can be seen from the figure, the superpixels boundaries trace the boundaries of the ionized regions extremely well.

In the remainder of this section we study the 21-cm PDFs generated from the superpixels. These are crucial for stitching superpixels together into ionized regions. We also investigate the under-segmentation error for different choices of the number of superpixels.

\subsection{21-cm PDFs}
\label{sec:limRes_PDF}
Above, in Fig.~\ref{fig:pdf_noise_nonoise} we have seen that the 21-cm PDF constructed directly from the pixels of the noisy data displays the signatures of reionization (asymmetric tail or bimodality) less clearly than the noise free data does. For example, during the early stages (\FIX{$x^\mathrm{v}_\mathrm{HII}= 0.2$}), the asymmetric tail is barely visible in the noisy data. The PDF generated from the superpixels will obviously be less noisy as the superpixels combine the data from several pixels. 

\begin{figure}
  \centering
  \includegraphics[width=0.48\textwidth]{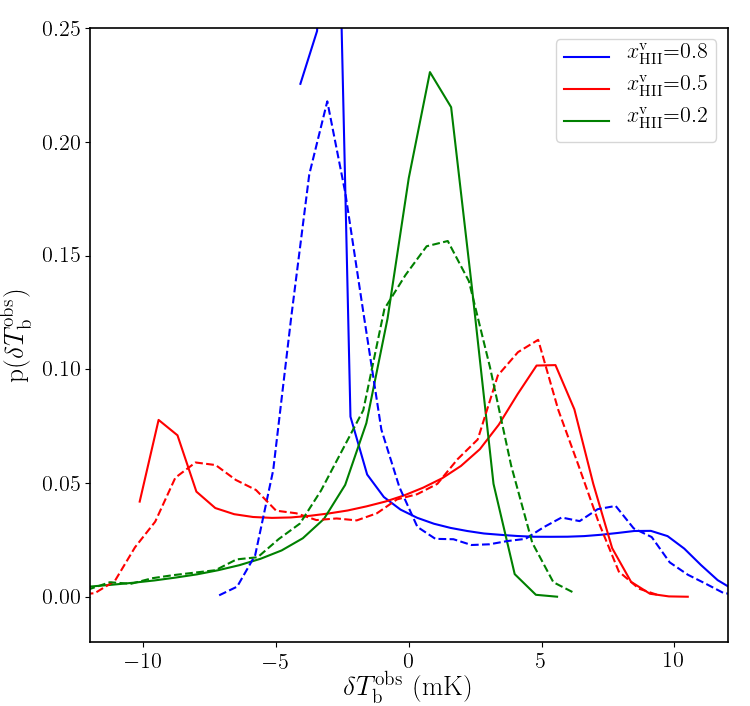}
  \caption{PDFs of the 21-cm signal for the three different stages of reionization. The solid line shows the single pixel PDF of the smoothed 21cm signal \FIX{without noise}. The dashed line shows the superpixel PDF of the same data \FIX{with noise}.}
  \label{fig:pdf_noise_slic}
\end{figure}

Fig.~\ref{fig:pdf_noise_slic} compares the \FIX{noiseless} single pixel PDFs (solid lines) with the superpixel PDFs (dashed lines) at different stages of reionization. \FIX{The superpixel PDF reproduces the main features of the noiseless single pixel PDF very well. Even though there are some differences in normalization, the general shape, bimodal or unimodal as well as the positions of the peaks show a good match. Definitely compared to the single pixel noisy PDF shown in Fig.~\ref{fig:pdf_noise_nonoise} the superpixel PDF does a much better job of recovering the noiseless single pixel PDF.} In our scheme these superpixel PDFs are used to label the superpixels as ionized or neutral after which they can be stitched together to find the boundaries of ionized regions as shown in Fig.~\ref{fig:superpixeled}.



\subsection{Number of superpixels}
\label{sec:seg_err_result}
\begin{figure}
  \includegraphics[width=0.46\textwidth]{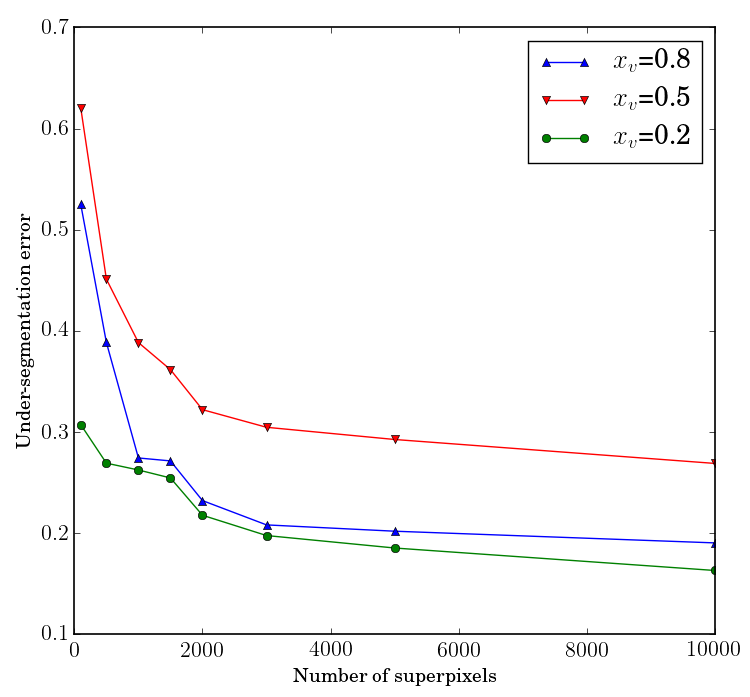}
  \caption{The under-segmentation error as a function of the number of superpixels for three different stages of reionization in our simulation. Beyond 3000 -- 5000 superpixels the under-segmentation error does not change much.}
  \label{fig:under_seg_err}
\end{figure}

Next we test the effect of the choice for the number of superpixels on the accuracy with which the ionized regions are identified. The under-segmentation error defined in Section~\ref{sec:seg_err} is used to quantify the error made while tracing the boundaries of ionized regions. We calculated the under-segmentation errors for three phases of reionization (\FIX{$x^\mathrm{v}_\mathrm{HII}$}= 0.2, 0.5 and 0.8) and for the number of superpixels ranging from 100 to 10,000. 

The curves in Fig.~\ref{fig:under_seg_err} show that increasing the number of superpixels decreases the under-segmentation error, but the improvement levels off after $\sim$5000 superpixels. Using a larger number of superpixels makes the segmentation computationally more expensive without any substantial improvement in the identification. If one takes two to three times the FWHM of the resolution as the smallest scale in Eq.~\ref{eq:ansatz}, the number of superpixels to use falls in the range $\sim 10,000$--3500. Therefore, we have used 5000 superpixels for all further over-segmentation calculations in this paper.


\section{Comparison of segmentation methods}
\label{sec:comparison}
In this section we compare the ionized regions identified by the three methods ({\lq Mean\rq}, {\lq K-Means\rq} and {\lq SLIC\rq}) using a correlation test and several global statistical measures. Where needed we use {\lq Ref-Bin\rq}, defined in Section~\ref{sec:performance} as a proxy for the true distribution of ionized regions.

\subsection{Correlation test}
\begin{figure}
  \includegraphics[width=0.48\textwidth]{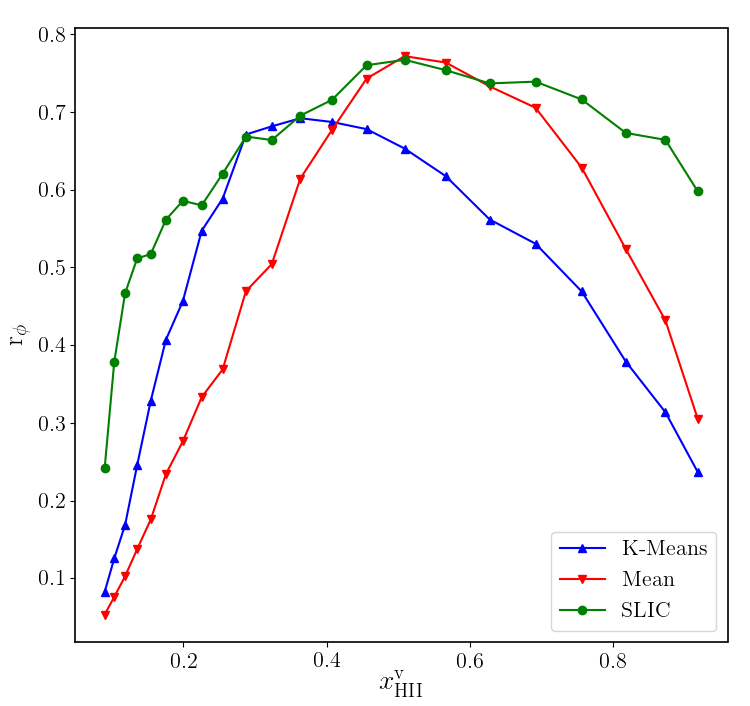}
  \caption{The correlation of the recovered binary fields with the Ref-Bin field for different segmentation methods at different reionization epochs. 
  }
  \label{fig:mcc_phi_binaries}
\end{figure}

To test how well the three methods recover the intrinsic shapes and sizes of the ionized regions, we correlate their results with the Ref-Bin data set. For this calculation we have constructed binary fields in which the pixels are labelled as ionized or not ionized. The correlation coefficient used is the phi coefficient \citep[$r_\mathrm{\phi}$, see][]{cramer1946methods}. There are just four combinations for the data in each pair of binary fields: (0,0), (0,1), (1,0), (1,1). The two binary fields are considered positively associated if most of the data are (0,0) and (1,1). The expression for r$_\mathrm{\phi}$ is given by
\begin{equation}
\mathrm{r}_\mathrm{\phi} = \frac{n_{11}n_{00}-n_{01}n_{10}}{\sqrt[]{n_{\cdot 0}n_{\cdot 1}n_{0 \cdot}n_{1 \cdot}}},
\end{equation}
where $n_{00}$, $n_{01}$, $n_{10}$ and $n_{11}$ are the number of the (0,0), (0,1), (1,0), (1,1) combinations found in the two fields and $n_{\cdot 0}$, $n_{\cdot 1}$, $n_{0 \cdot}$ 
and $n_{1 \cdot}$ are the number of 0's and 1's in each of the fields respectively. 

Figure~\ref{fig:mcc_phi_binaries} shows the $r_{\phi}$ calculated for the recovered binary fields from the three methods: Mean, K-Means, and SLIC. 
Even though the Mean method performs well at certain epochs, the $r_{\phi}$ varies too much over redshift for this to be a reliable method. When analysing real data we would not necessarily know if we are in the regime where this methods works best. Also it is a poor method when we want to compare properties of ionized bubbles from different epochs. The same holds true for the K-Means results.
Therefore, SLIC is the most robust method to apply on the 21-cm observation when there is no prior knowledge about the reionization history.

\FIX{All methods show a decrease in performance for low ionization fractions. This is due to several factors. First of all the noise level increases for higher redshifts. Secondly, a larger fraction of the ionized regions is in the form of small HII regions which are only just above the resolution limit of the observations. The latter factor is made worse by the fact that the resolution decreases for higher redshifts. A third and related factor for SLIC and K-Means is that below $x^\mathrm{v}_\mathrm{HII}=0.2$ the PDF becomes unimodal so the less accurate maximum deviation method is used to define the threshold. Lastly, at low ionization fractions the Mean method will misidentify low density regions as ionized ones, as was shown in \citet{2017MNRAS.471.1936K}.}

\subsection{Statistics}
After identifying the ionized regions, we want to extract information about the EoR. 
In this section, we show how the statistics from the binary fields created by the three segmentation methods introduced in Section~\ref{sec:bub_seg} compare with the statistics from the true ionized regions. We again use the Ref-Bin at different epochs as the proxy for the true ionized regions. The statistical quantities compared here are the power spectra of the binary field and the Bubble Size Distributions (BSDs). The power spectra will inform us how well the identification methods work at different scales. The BSDs show the potential of these methods to identify regions of different sizes.



\subsubsection{Power spectra}
\label{sec:PS}
\begin{figure*}
  \centering
  \includegraphics[width=0.95\textwidth]{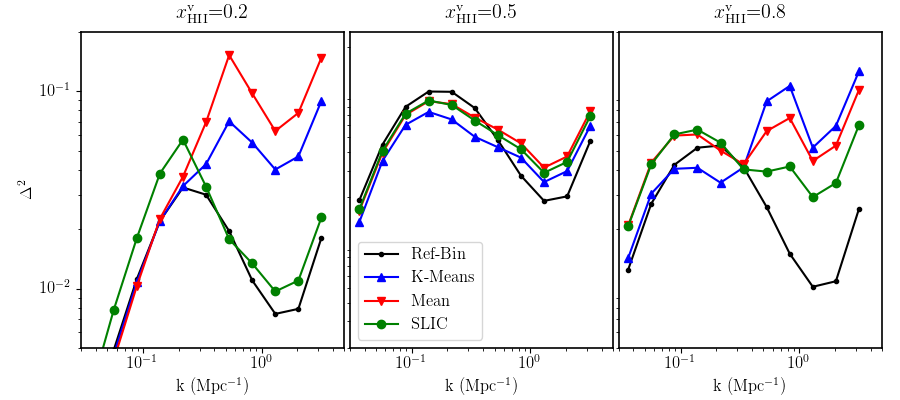}
  \caption{The dimensionless power spectra of the binary ionization fields  identified by the three methods for the early, middle and late stages of reionization.}
  \label{fig:ps_fields}
\end{figure*}

The spherically averaged power spectrum $P_{xx}(k)$ of the ionization fraction field can be defined as
\begin{equation}
\langle \hat{x}(\mathrm{\bf k}) \hat{x}^{\star}(\mathbf{k'})\rangle = (2 \pi)^3 \delta_D(\mathbf{k - k'}) P_{xx}(k),
\end{equation}
where $\hat{x}(\mathrm{\bf k})$ represents the Fourier components of the binary ionization field $x$. We present the results in terms of the dimensionless power spectrum of the ionization field $\Delta^2_{xx}=k^3P_{xx}(k)/2\pi^2$.
We would like to point out that the power spectrum of the binary ionization field is not the same as the power spectrum of the real ionization fraction. The physical interpretation of the binary ionization fields requires more work which is beyond the scope of the paper. Here we only use it to compare the different segmentation methods.  

The scale dependence of $\Delta^2_{xx}$ for the three different segmentation methods considered in this work is shown in Fig.~\ref{fig:ps_fields} at three different stages of reionization. Around the mid-point of reionization, all three methods perform quite well. However, both at the early and late stages quite large deviations from the reference power spectrum are seen and these are larger for the Mean and K-Means methods. The former performs the worst during the earlier stages and the latter during the later stages. SLIC performs best in all cases with a result that is very close to the reference power spectrum for the earlier and middle stages and slightly off for the later stages. \FIX{This is in contrast with the correlation results from the previous section. However, it should be realised that the cross-correlation measures something different from the power spectra. First of all the correlation coefficient depends on both the ionized and neutral regions, whereas the power spectrum calculated here only depends on the ionized regions. Furthermore the power spectra mostly differ for small scales and show better agreement at large scales. The latter will dominate the correlation coefficient.}

\subsubsection{Bubble size distribution (BSD)}
\begin{figure*}
  \centering
  \includegraphics[width=0.95\textwidth]{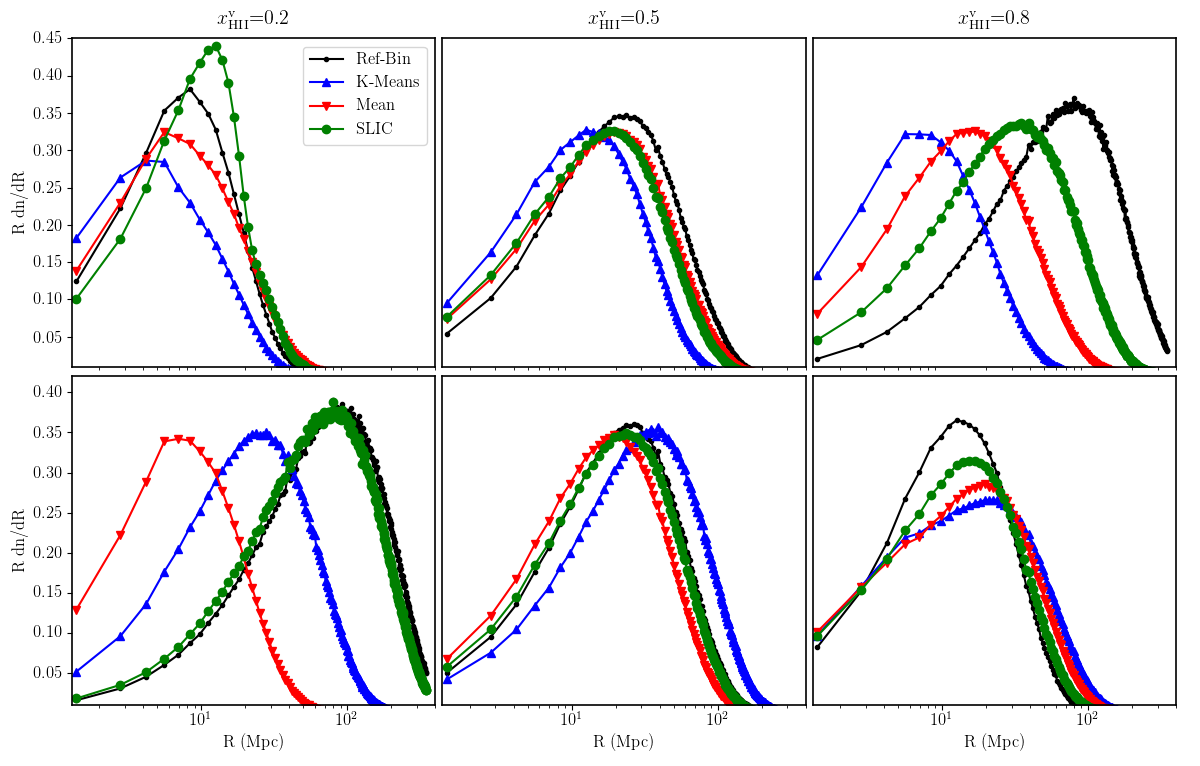}
  \caption{The MFP-BSDs identified by the three segmentation methods for the three different stages of reionization. The top panels show the BSDs of ionized regions whereas the bottom panels display the BSDs of neutral (non-ionized) regions.}
  \label{fig:mfp_sizes}
\end{figure*}
BSDs have been studied in great depth in previous papers. Originally they were introduced as tools to analyse and compare EoR simulations \citep[e.g.][]{2011MNRAS.413.1353F,2016MNRAS.461.3361L}. More recently, \citet{2017MNRAS.471.1936K} and \citet{giri2017bubble} have studied BSDs derived from 21-cm data sets. 

Due to the complexity of the shapes of ionized regions, there is no unique way to define their sizes. Hence a range BSD methods have been developed. Here, we use the so-called mean-free-path (MFP) method  to compare the performance of our three segmentation methods \citep{2007ApJ...669..663M}. The MFP method starts with choosing a random point inside an ionized region. From this point a ray is traced in a random direction until it hits the boundary of the ionized region\footnote{The boundary is defined as the last ionized pixel before it hits a neutral pixel.} and the resulting length of this ray is recorded. This is a Monte Carlo method for determining the size distribution of ionized bubbles. 
The distribution of the recorded lengths of the rays converges to the BSD when the process is repeated numerous times. We have used 10$^7$ rays. 

We compare the MFP-BSDs at 20 per cent, 50 per cent and 80 per cent global ionization stages in Fig.~\ref{fig:mfp_sizes}. In the top panels, the sizes of the ionized regions are compared whereas in the bottom panels, we present the results for the neutral regions.
The neutral regions studied here are defined as the complimentary part of the ionized region, which means that they are the regions which were not labelled as ionized. Although this is correct for the Mean and K-Means methods, for SLIC it may be better to employ a different stitching algorithm when the goal is to identify neutral regions. In the current segmentation the BSDs of the neutral regions could be affected by superpixels of mixed content and noisy cells which were not labeled as ionized. 

We observe that around the midpoint of reionization all methods give similar BSDs which lie close to that of the reference data set. This agrees with the power spectra results in Section~\ref{sec:PS}. For the early stages, for the ionized regions the BSDs for all methods differ but at least peak at a size close to that of the reference data. The BSD from SLIC has fewer small regions, probably because they are too small to be picked up by the superpixels. The Mean method has more larger regions than the reference set, and K-Means has more small regions and fewer large regions than the reference set. For the neutral regions, the BSD of regions identified by SLIC is the only one close to the reference set.

At the later stages of reionization, the ionized regions BSDs from all the methods peak at lower values than the reference set. However, the SLIC result is much closer to the reference than the others. 
In case of the neutral binary fields, all the BSDs are fairly close to each other. All methods overestimate the relative number of large regions and underestimate the relative number of small regions. Of the three methods, SLIC again is closest to the reference set.

From these results we conclude that the BSDs of the regions identified by SLIC are more reliable compared to the other methods. This makes SLIC a better choice for studying the evolution of bubble sizes during the EoR. Here we have only shown results from one BSD algorithm. However, we can report that SLIC performs better also for other BSD algorithms. 

\subsection{Mean ionized fraction}
\label{sec:mean_ionization_vol}
\begin{figure}
  \centering
  \includegraphics[width=0.45\textwidth]{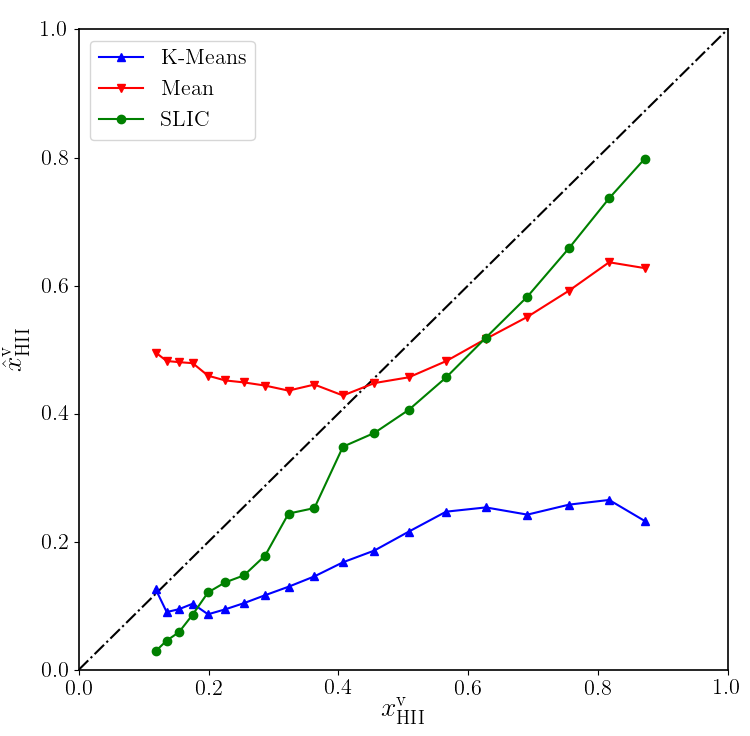}
  \caption{The evolution of the mean ionization fraction by volume $\hat{x}^\mathrm{v}_\mathrm{HII}$ measured from the results of the three segmentation methods \FIX{as a function of $\hat{x}^\mathrm{v}_\mathrm{HII}$ estimated from the binary fields.}}
  \label{fig:x_volume_estimate}
\end{figure}

From the binary fields, we can easily calculate the fraction of the volume that is ionized, \FIX{$x^\mathrm{v}_\mathrm{HII}$}. From this we can obtain an estimate of how far reionization has progressed. 
Fig.~\ref{fig:x_volume_estimate} shows the measured ionization fractions against the true fraction for the three segmentation methods. It is obvious that the Mean method gives a very unphysical estimation of the mean ionized volume. The estimated values \FIX{$\hat{x}^\mathrm{v}_\mathrm{HII}$} remain very close to 50 per cent throughout the reionization history and the curve does not show a consistent positive slope. This method clearly cannot be used to estimate \FIX{$\hat{x}^\mathrm{v}_\mathrm{HII}$}.

The K-Means method underestimates \FIX{$x^\mathrm{v}_\mathrm{HII}$} but the evolution has the correct slope between \FIX{$x^\mathrm{v}_\mathrm{HII}$}=0.2 and \FIX{0.6}. The corresponding result for noiseless data was shown in \citet{giri2017bubble}, fig.~3. There the \FIX{$\hat{x}^\mathrm{v}_\mathrm{HII}$} results below \FIX{$x^\mathrm{v}_\mathrm{HII}$}=0.1 were noisy. In the current results, this is no longer the case which is due to the introduction of the maximum deviation algorithm for when the PDF is unimodal. However, those measurements show a decreasing \FIX{$\hat{x}^\mathrm{v}_\mathrm{HII}$} for increasing \FIX{$x^\mathrm{v}_\mathrm{HII}$} indicating that K-Means should not be used for estimating the mean ionization fraction during early reionization. We also see that the evolution of \FIX{$\hat{x}^\mathrm{v}_\mathrm{HII}$} levels off for \FIX{$x^\mathrm{v}_\mathrm{HII} > 0.6$}. As this behaviour was not seen in \citet{giri2017bubble}, this has to be due to the presence of noise.

Although the SLIC results also underestimate $x^\mathrm{v}_\mathrm{HII}$, the evolution has the correct slope during the EoR and the difference with the true value is never more than 0.1. Also for this metric, SLIC clearly outperforms both K-Means and Mean.

\citet{giri2017bubble} already showed that measurements of $x^\mathrm{v}_\mathrm{HII}$ from observations will always underestimate the true values. This is due to the fact that the smallest ionized regions are not resolved in the 21-cm data. However, by taking this into account, measurements of \FIX{$\hat{x}^\mathrm{v}_\mathrm{HII}$} will still be useful for tracing the evolution of reionization.

\section{Further applications of superpixels}
\label{sec:further}
In the previous section, we have seen that segmenting the 21-cm data sets using superpixels is superior to the other segmentation approaches in terms of recovering the power spectrum of the ionization fraction field, the BSDs of ionized
regions and the mean ionization fraction by volume. However, segmenting the data into superpixels allows us to extract other physical information from the 21-cm data. This section describes a few examples of further application of superpixels, namely estimates of the global 21-cm signal, PDF measurements, as well as an estimate of the number of ionizing photons produced within each superpixel.

\subsection{Global signal or mean ionization by mass}
\label{sec:mean_ionization_mass}
\begin{figure}
  \centering
  \includegraphics[width=0.485\textwidth]{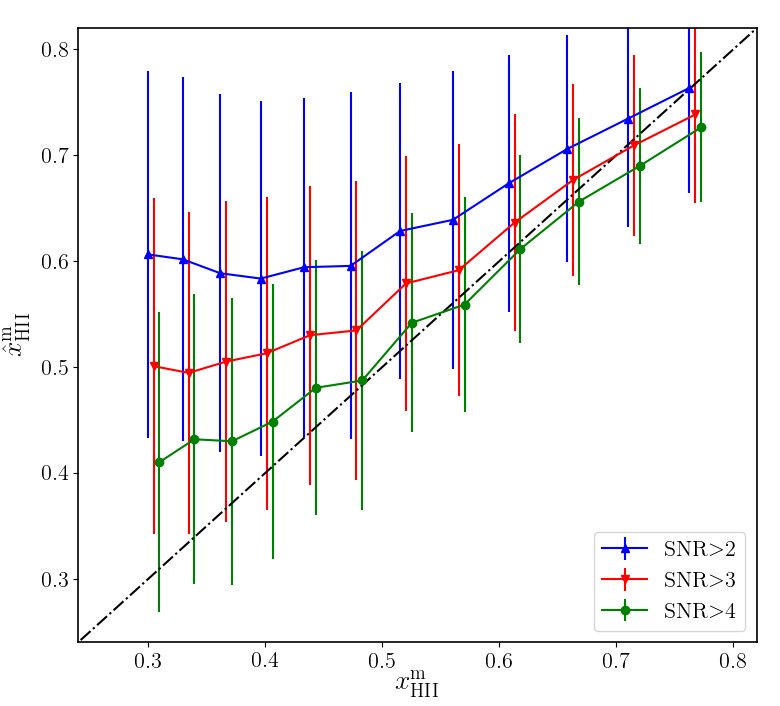}
  \caption{The evolution of the mean ionization fraction (by mass) as measured from the 21-cm signal of highly ionized superpixels. We use the SNR of superpixels to select the highly ionized ones where higher a SNR implies a more reliable identification. The points for SNR $>$ 2, 3 and 4 have been shifted by a tiny amount to make the error-bars clearly visible.}
  \label{fig:x_mass_estimate}
\end{figure}

The absence of zero-length baselines in interferometric observations deprives us from measuring the absolute value of the 21-cm brightness temperature, the average of which, \FIX{$\langle\delta T^\mathrm{sim}_\mathrm{b}\rangle$} is known as the global signal. The measured interferometric data will always have a mean signal of zero. In this section, we propose a method to estimate \FIX{$\langle\delta T^\mathrm{sim}_\mathrm{b}\rangle$} using the superpixels. 

The intrinsic 21-cm signal from the ionized regions is zero. However, an interferometer such as SKA1-Low will instead see a negative signal whose absolute value is equal to the mean signal in that frequency channel. \citet{2015aska.confE..10M} suggested that the global signal can therefore be estimated by measuring the brightness temperature from the well resolved portions of ionized bubbles. 
\FIX{In the high spin temperature limit, the observed signal is related to the true or simulated signal in the following way,}

\begin{equation}
\delta T^\mathrm{obs}_\mathrm{b} = \delta T^\mathrm{sim}_\mathrm{b} - \langle \delta T^\mathrm{sim}_\mathrm{b} \rangle
\label{eq:observed_Tb}
\end{equation}

\noindent \FIX{As there will be no signal from the fully ionized pixels ($\delta T^\mathrm{sim}_\mathrm{b}=0$), the observed signal from those will be associated to the global signal of the observed FOV.}

The segmentation with superpixels allows us to select and analyse such well resolved ionized regions. They should have a constant 21-cm signal together with noise. Less well resolved ionized regions will have a mix of 21-cm fluctuations and noise. Therefore the superpixels with the lowest signal and the lowest variance are good candidates for determining the global signal. The absolute value of the mean signal from all these superpixels will provide an estimate for \FIX{$\langle\delta T^\mathrm{sim}_\mathrm{b}\rangle$}. 

Although the mean global 21-cm signal is an interesting quantity in its own right, we choose to represent it here by the mean ionization fraction by mass, \FIX{$x^\mathrm{m}_\mathrm{HII}$}. As can be seen from Equation \ref{eq:dTb}, for the high spin temperature limit
\begin{eqnarray}
1-x^\mathrm{m}_\mathrm{HII} = \langle x_\mathrm{HI}(1+\delta)\rangle \nonumber \\ 
\propto \langle\delta T^\mathrm{sim}_\mathrm{b}\rangle,
\label{eq:x_m}
\end{eqnarray}
where $\langle.\rangle$ represents the average of that quantity. In other words for the high spin temperature limit and a given cosmology \FIX{$\langle\delta T^\mathrm{sim}_\mathrm{b}\rangle$} and \FIXII{1-$x^\mathrm{m}_\mathrm{HII}$} are fully equivalent.

In Fig.~\ref{fig:x_mass_estimate}, the measured \FIX{$\hat{x}^\mathrm{m}_\mathrm{HII}$} considering different SNR cutoffs have been plotted. The SNR here is the mean signal of the superpixel over its rms value. The error bars show the variation in the measured values. 
We can see the error bar getting smaller when superpixels with better SNRs are selected. It is also seen that the estimate is better during the later stages of reionization. In those later stages, the number of fully ionized superpixels is larger. \FIX{When selecting the highest SNR superpixels (SNR $>4$) the true value is
always within the error bars and the estimate is always within 0.1 of the true value.}

We have to caution that this method relies on a proper calibration of the 21-cm signal and the absence of any residual foreground signals which could add additional signal to the data cubes. Studies with more realistic data representations including residual foregrounds are needed to establish whether these will pose a problem for this measurement.

\subsection{PDF measurements}
\citet{2010MNRAS.406.2521I} have shown that the progress of reionization can be studied using the 21-cm PDF. However, as we saw above, system noise can affect the PDFs in a substantial way (Fig.~\ref{fig:pdf_noise_nonoise}) and therefore the reionization history derived from noisy PDFs may be biased (Fig.~\ref{fig:pdf_noise_nonoise}). The PDFs measured from the superpixels are less affected by noise (see Fig.~\ref{fig:pdf_noise_slic}) and can therefore provide a better way to perform similar studies. 

The less noisy superpixel PDFs also can improve the measurement of higher order one-point statistics such as the skewness and kurtosis. These characterisations of the non-Gaussian nature of the PDF have been shown to be useful statistics for the 21-cm signal \citep[e.g.][]{2009MNRAS.393.1449H,2014MNRAS.443.3090W} but are also difficult to extract from noisy data.

Since the superpixels are not all of equal size, the noise reduction and smoothing they introduce is not uniform. Therefore the superpixel PDF and quantities derived from it such as the skewness and kurtosis, may have a complicated relation to the intrinsic PDF. This may introduce biases which should be understood before using the superpixel PDF instead of the single pixel PDF. We postpone an in depth investigation of the usefulness of the superpixel PDFs to a future paper.

\subsection{Source properties}

\begin{figure*}
  \centering
  \includegraphics[width=0.99\textwidth]{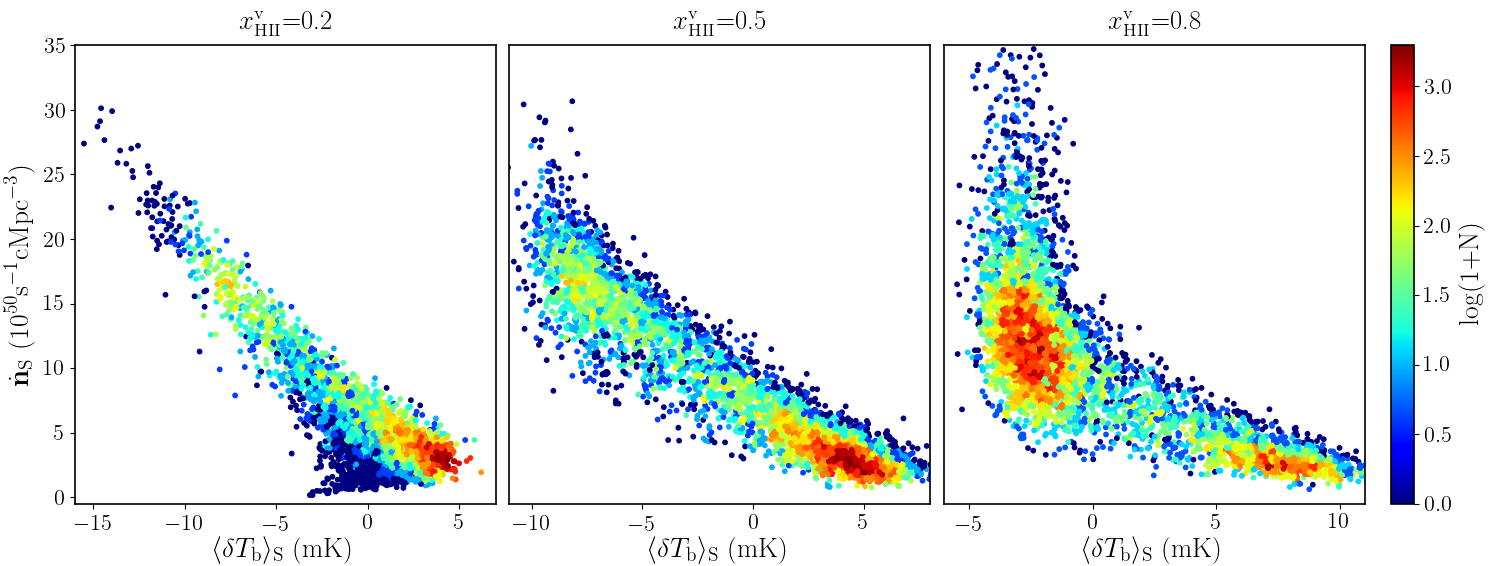}
  \caption{Scatter plots of the photons production rate (per unit volume) inside superpixels against the mean of signal of the superpixel. The different panels shows different epochs of reionization (\FIX{$x^\mathrm{v}_\mathrm{HII}$} = 0.2, 0.5, 0.8). The colour of the points plotted gives the number density (N) of the number of superpixels.}
  \label{fig:scatter_nphotonperpixel}
\end{figure*}

One of the main aims of studying the reionization process is to establish the properties of the sources of reionization. Previous papers have shown that the statistical properties of the 21-cm signal and their evolution depend on the source properties and can be used to constrain the nature of these sources \citep[e.g.][]{2015MNRAS.449.4246G,2017MNRAS.468.3869S}.


Locally, the size and shape of an ionized region is determined by the positions of the sources of ionizing photons, their photon fluxes and the history of these. Tomographic images of the 21-cm signal as will be produced by SKA1-Low will allow us to characterize these shapes and sizes which could provide additional information about the sources beyond the statistical ones. 
The images will allow us to pinpoint regions in the sky which based on the presence of a large ionized region should be rich in high redshift galaxies and would be excellent targets for deep optical/infrared surveys.

However, could we do more with the 21-cm tomographic images? Our simulation results inform us of the rate of ionizing photons produced in different regions. In Fig.~\ref{fig:scatter_nphotonperpixel}, we show scatter plots of the summed rates of all sources located within a superpixel $\dot{n}_\mathrm{S}$ against the average value of the 21-cm signal of that superpixel, $\langle \delta T_\mathrm{b} \rangle_\mathrm{S}$. We show this for three stages of reionization, namely at \FIX{volume averaged} ionization fractions of 0.2, 0.5 and 0.8. It is clear that there exists a \FIX{tight} anti-correlation between the local photon production rate $\dot{n}_\mathrm{S}$ and the value of the 21-cm signal \FIX{which allows us to estimate the local photon production rate directly from the value of $\langle \delta T_\mathrm{b} \rangle_\mathrm{S}$.}

\FIX{In two cases we see a larger spread in photon production rates for a given value of $\langle \delta T_\mathrm{b} \rangle_\mathrm{S}$. In the left panel for the range $-5<\langle \delta T_\mathrm{b} \rangle_\mathrm{S}<0$~mK, photon production rates vary from 0 to $\sim 10^{51}$~s$^{-1}$~Mpc$^{-3}$. Superpixels in this range of $\langle T_\mathrm{b} \rangle_\mathrm{S}$ consist of a mix of low density neutral regions and (partially) ionized higher density regions. Note however that the number of superpixels with low photon production range are relatively rare. }

\FIX{The other case is for $x_\mathrm{HII}^\mathrm{m}=0.8$ (right panel) when most of the Universe has already reionized. A large range of photon production rates are found for the range $-5<\langle \delta T_\mathrm{b} \rangle_\mathrm{S}<0$~mK. This range now corresponds to ionized regions and obviously once a region has reionized the correlation with the current photon production rate is lost. However, at other values of the 21-cm signal and around the midpoint of reionization we find a tight relation between $\langle \delta T_\mathrm{b} \rangle_\mathrm{S}$ and $\dot{n}_\mathrm{S}$ which allows us to estimate the local photon production rate directly from the value of $\langle \delta T_\mathrm{b} \rangle_\mathrm{S}$.}

These results indicate that 21-cm tomography may be a fairly good tracer of the instantaneous photon production rate. Clearly this anti-correlation should be investigated further for many different reionization models to establish whether there is an absolute correlation between the 21-cm signal and the local photon production rate. If this is the case, the 21-cm tomography would not only allow us to select interesting regions for galaxy surveys but also to connect the galaxy properties to the ionizing photon production rate, potentially constraining the fraction of ionizing photons escaping from a galaxy population. The segmentation provided by the superpixel method clearly enables this kind of analysis. 


\section{Summary and discussion}
In this study, we have introduced the superpixel method to identify the ionized regions in noisy 21-cm tomographic data. This method is based on over-segmentation of the data sets using the SLIC algorithm. Using a wide range of tests we have shown that this method performs better and is more robust compared to other single pixel (intensity-based) methods that have been previously used. After identifying ionized bubbles these can be analysed in different ways, for example by considering their size distribution using different size metrics \citep{giri2017bubble}. Identification of ionized regions may also allow a measurement of the global 21-cm signal in interferometric data. Furthermore, it enables estimates of the production of ionizing photons in different regions.



Since the superpixel method effectively reduces the resolution at which the data is analysed, its expected performance will be worse during the early stages of reionization when there are many small ionized regions present. However, it is not clear that any other segmentation method will perform better as separating density fluctuations, ionization fluctuations and noise will be intrinsically difficult during this stage. As long as ionized regions are isolated and regular in shape, the matched filtering technique as first explored by \citet{2007MNRAS.382..809D} and more recently by \citet{2016MNRAS.460..827G}, can be used to target these early phases. However, as shown by \citet{2016MNRAS.457.1813F}, the ionized regions are expected to start to percolate into one large connected cluster already around a global ionization fraction of 0.1, so the phase of isolated ionized bubbles will likely be quite short.

The research field of image segmentation has produced a wide range of methods to find structures in images. Based on the properties of the 21-cm data sets expected from SKA1-Low, we expect the superpixel method to be among the better ones for identifying ionized regions but we cannot exclude that other region-based methods will perform equally well. In the presence of noise, methods based on single pixel intensities will always struggle. Of course it is possible to reduce the noise by further reducing the resolution of the data. This is likely to diminish the difference in performance between superpixels and the single pixel methods although at the cost of losing smaller scale features. Here we limited ourselves to studying one resolution, defined by a maximum baseline of 2~km, and one noise level, defined by an integration time of 1000 h. We will explore a wider range of cases as well as the trade off between noise levels and resolution in a future paper (Mellema et al., in prep).

For our comparison of segmentation methods, we considered only one reionization scenario. Obviously, the real Universe may have reionized in a very different way so exploring a wider range of scenarios would be good. However, for investigating image segmentation techniques, the actual reionization scenario is likely to be less important, unless it follows a completely different geometry, such as a featureless ionization fraction field in which case there will be no ionized regions to identify any way. For different scenarios of patchy reionization, the quantitative results will differ from the ones we presented here, but it would require very contrived scenarios for the superpixel method to perform worse than the other segmentation methods.

A larger worry is the impact of spin temperature fluctuations which we have ignored here. As already explained in \citet{giri2017bubble}, spin temperature fluctuations impose fundamental difficulties for identifying ionized regions as they modify the PDF of 21-cm values. Without clear PDF features that can be associated with ionized regions, it will be difficult to identify them, especially since spin temperature values close to $T_\mathrm{CMB}$ produce a signal value similar to that of ionized regions. Whether we will be able to identify ionized regions will therefore depend on the level and structure of the spin temperature fluctuations and the assumptions for the properties of the X-ray sources heating the neutral IGM. A first step for investigating the impact of spin temperature fluctuations will be an exploration of the PDFs of 21-cm values for a wide range of scenarios and redshifts.
When constructing our 3D 21-cm data sets we ignored a number of line of sight effects which the real data will have, namely the light cone effect \citep{2012MNRAS.424.1877D,2014MNRAS.442.1491D} and redshift space distortions \citep{2012MNRAS.422..926M, 2013MNRAS.435..460J}.  Although these effects can impact for example both power spectra and BSDs \citep[e.g.][]{giri2017bubble}, they are unlikely to impact the superpixel method. SLIC looks for local clustering of points using a modified distance metric. The scale at which the algorithm runs is defined by the number of superpixels that we choose. As can be seen in Fig.~\ref{fig:under_seg_err}, the number of the superpixels used is several 1000s. Therefore, the extent of the superpixels in the frequency direction will be small enough for the algorithm not to influenced by these line of sight effects.

Another effect which creates anisotropy between the frequency and angular directions are residuals after foreground subtraction. These have to be limited to large scale variations so as not to destroy any measurements of the 21-cm signal and therefore will also not impact the SLIC algorithm. However, they may impact the values of the signal found in the superpixels and therefore affect the other applications from Section~\ref{sec:further} which depend on these values. The light cone effect and redshift space distortions may of course also impact the values but these effects can be calibrated through simulations.

The capability of the future SKA1-Low to image the 21-cm signal from the EoR tomographically will enable us to answer fundamental questions about the reionization process, the sources of ionizing photons and cosmology. To find these answers requires advanced image processing methods such as the one described in this paper. These will allow us to extract statistical information about the sizes and shapes of different types of regions. Here we presented one stitching algorithm which identifies ionized superpixels. The properties of these ionized regions can then be analysed statistically. Other stitching algorithms may be designed to identify neutral regions or, in the case of spin temperature fluctuations, cold or hot regions. The statistical properties of the superpixels found may themselves contain information about the reionization process. The further development of such image processing algorithms will be important for the future of the new research field of 21-cm cosmology.

\section*{Acknowledgements}
This work was supported by Swedish Research Council grant 2016-03581. We acknowledge that the results in this paper have been achieved using the PRACE Research Infrastructure resources Curie based at the Très Grand Centre de Calcul (TGCC) operated by CEA near Paris, France and Marenostrum based in the Barcelona Supercomputing Center, Spain. Time on these resources was awarded by PRACE under PRACE4LOFAR grants 2012061089 and 2014102339 as well as under the Multi-scale Reionization grants 2014102281 and 2015122822. Some of the numerical computations were done on the Apollo cluster at The University of Sussex as well as on resources provided by the Swedish National Infrastructure for Computing (SNIC) at PDC, Royal Institute of Technology, Stockholm. We thank Ilian Iliev and Bart Pindor for useful discussions.






\bibliography{references}



\appendix


\bsp	
\label{lastpage}

\end{document}